\documentclass[%
 aps,%
 amsmath,amssymb,superscriptaddress,
 reprint,%
]{revtex4-1}

\usepackage{graphicx}% Include figure files
\usepackage{dcolumn}% Align table columns on decimal point
\usepackage{bm}% bold math
\usepackage{hyperref}
\usepackage{tabularx}
\begin{document}

\preprint{AIP/123-QED}

\title{High phonon-limited mobility of charged and neutral excitons in mono- and bilayer MoTe$_2$}% Force line breaks with \\

\author{Sophia Helmrich}
\email{sophia.helmrich@tu-berlin.de}
%\phone{+49 (0)30 314 23636}
\author{Alexander W. Achtstein}
\author{Hery Ahmad}
\author{Matthias Kunz}
\affiliation{Institut f\"ur Optik und Atomare Physik, Technische Universit\"at Berlin, Germany}
\author{Bastian Herzog}
\affiliation{Institut f\"ur Optik und Atomare Physik, Technische Universit\"at Berlin, Germany}
\author{Oliver Sch\"ops}
\affiliation{Institut f\"ur Optik und Atomare Physik, Technische Universit\"at Berlin, Germany}
\author{Ulrike Woggon}
\affiliation{Institut f\"ur Optik und Atomare Physik, Technische Universit\"at Berlin, Germany}
\author{Nina Owschimikow}
\email{nina.owschimikow@physik.tu-berlin.de}
\affiliation{Institut f\"ur Optik und Atomare Physik, Technische Universit\"at Berlin, Germany}
\affiliation{Ferdinand-Braun-Institut f\"ur H\"ochstfrequenztechnik, Berlin, Germany}

\date{\today}% It is always \today, today,

\begin{abstract}

We analyze the lineshape of the quasiparticle photoluminescence of monolayer and bilayer molybdenum ditelluride in temperature- and excitation intensity-dependent experiments. We confirm the existence of a negatively charged trion in the bilayer based on its emission characteristics and find hints for a coexistence of intra- and interlayer trions with a few meV splitting in energy. From the lineshape analysis of exciton and trion emission we extract values for exciton and trion deformation potentials as well as acoustical and optical phonon-limited mobilities in MoTe$_2$, with the mobilities showing the highest values so far reported for transition metal dichalcogenides.
\end{abstract}

\pacs{Valid PACS appear here}% PACS, the Physics and Astronomy
%Use showkeys class option if keyword
\maketitle

The growing demand for touchable and wearable electronic devices has sparked a great amount of research activity in the direction of flat and flexible electronically active materials. The most prominent example is the two-dimensional (2D) material graphene, which is nowadays readily manufactured and features a high carrier mobility \cite{Bolotin, Bruzzone} as an essential prerequisite for efficiently functioning devices. A drawback of graphene, however, is the lack of a finite direct bandgap, which limits its applicability in opto-electronics. Black phosphorous is another 2D material, which displays a high carrier mobility \cite{Bhaskar} and a direct bandgap tunable by layer thickness, but it is unstable and poisonous. Transition metal dichalcogenides (TMDCs) are graphene-like 2D materials, for which a finite direct bandgap emerges in the limit of a monolayer (ML). They are relatively robust and show promising results in experiments aimed at growing them in scalable processes like chemical vapor deposition \cite{WangLuo}. Carrier mobility in TMDCs, however, is an issue that needs to be addressed, as the amount of research in this direction is limited, and the existing results point at possible difficulties to be overcome. For molybdenum disulfide (MoS$_2$) a carrier mobility of 200-400~cm$^2$/Vs has been reported \cite{Kaasbjerg}, which clearly does not meet the requirements for real world applications. Apart from MoS$_2$, emitting in the visible spectral range \cite{MakLee2010}, %cite{Splendiani}
 there is a large TMDC family, in which the combination of molybdenum with a chalcogenide generally yields a bright exciton as the lowest energy excitonic level. Except for MoS$_2$, the carrier mobility has not been determined quantitatively. In particular molybdenum ditelluride (MoTe$_2$), which emits in the technologically relevant near infrared wavelength region \cite{Ruppert,Helmrich}, is a promising candidate for opto-electronic devices due to a high calculated (acoustic phonon limited) electron mobility of 2500~(10.000)~cm$^2$/Vs \cite{ZhangHuang,Huang2016}.

\par In this contribution, we study the photoluminescence (PL) of neutral excitons ($X^0$) and charged excitons (trions, $X^T$) in MoTe$_2$, and analyze the observed lineshape. We obtain exciton relaxation times as well as phonon coupling parameters and acoustical and optical phonon-limited quasiparticle mobilities. The investigations are carried out for ML as well as for bilayer (BL) MoTe$_2$, as the peculiar band structure of this material leads to the same band alignment in ML and BL \cite{Helmrich}. The MoTe$_2$ BL displays a bright emission twice as intense as the ML. In particular, this greatly facilitates the study of the trion, which has been so far quantitatively analyzed only in ML TMDCs \cite{Yang, Gao}, largely due to the BL of most TMDCs being dark due to the indirect bandgap.

\begin{figure}
\includegraphics[width=0.42\textwidth]{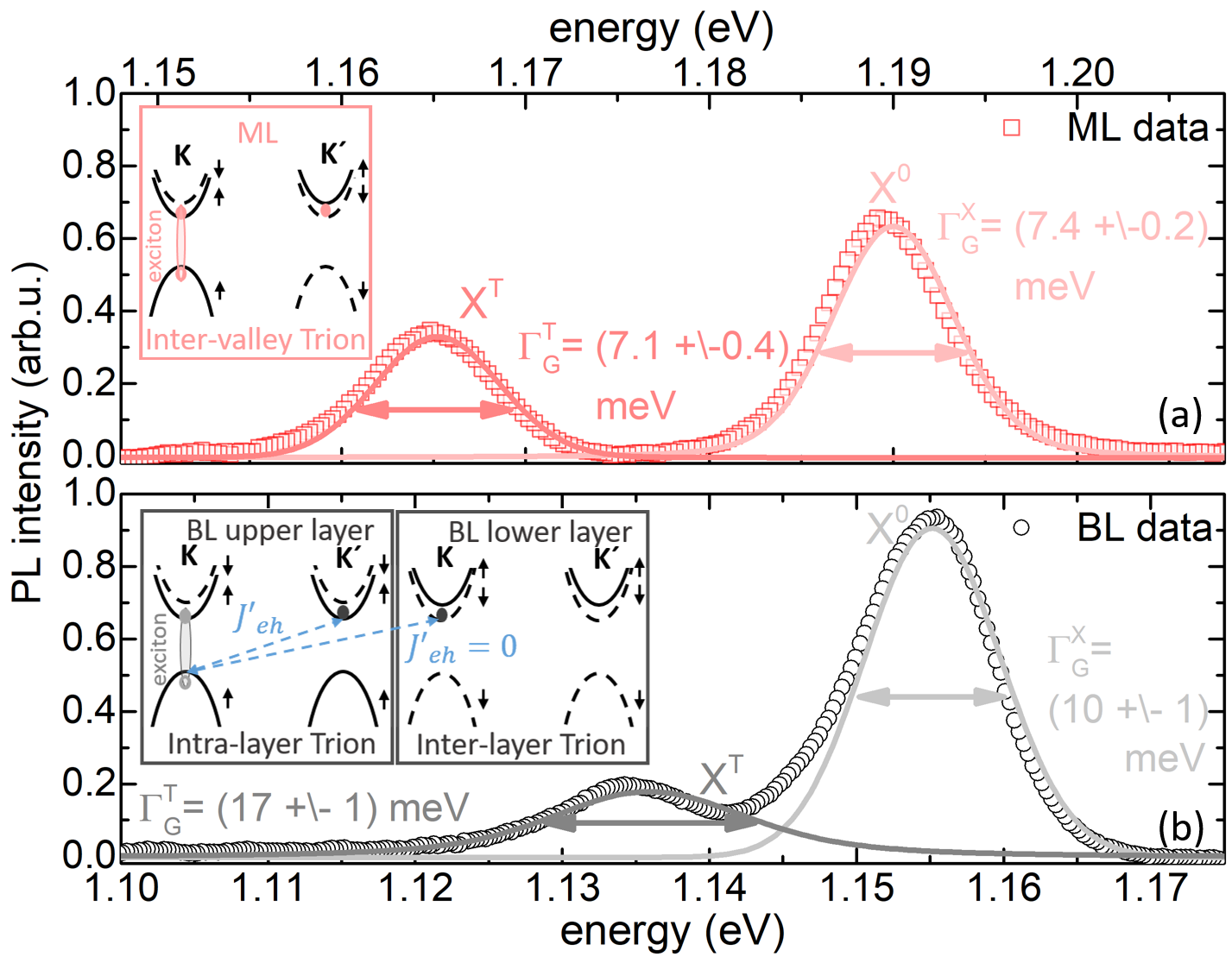}
\caption{(color online) Normalized PL spectra and Gaussian FWHM ($\Gamma_G$) for ML (a) and BL (b) exciton ($X^0$) and trion ($X^T$). (Inset) Schematics of the ML and BL K/K$'$ valleys plus possible excitonic and trionic states. $J'_{eh}$ denotes the exchange interaction energy between the excess electron and the electron-hole pair.}
\label{Gauss}
\end{figure}

\par We analyze the PL of a mechanically exfoliated flake of 2H MoTe$_2$ on a Si/SiO$_2$ ($80\, $nm) substrate with well separated ML and BL. For experimental details we refer to the SI. 
PL spectra for ML and BL MoTe$_2$ for a temperature of 10~K, and an excitation intensity of 35~W/cm$^2$ are shown in Fig.~\ref{Gauss}. To clearly display the lineshape of in particular the low energy peak, the spectra have been corrected for an impurity peak on the red side, for raw data and details of the correction procedure please refere to the SI. We tentatively denote the lower energy peak as the "trion" peak. Though the existence of a trion in the MoTe$_2$ BL has not yet been fully convincingly established in the literature, we will show that the feature observed by us behaves in accordance with assuming a negatively charged trion nature. The excitonic emission of ML and BL is well described by fitting with set of two Voigtian functions for exciton and trion. From the fits, we determine the central energies extrapolated to zero temperature of the ML excitonic ($X^0_{ML}$) and trionic ($X^T_{ML}$) peaks as $1.189\, $eV and $1.165\,$eV, respectively, resulting in a binding energy $E_{bind}$ of $(24.9\pm 0.1)\,$meV. The BL also displays two well-resolved peaks extrapolated to $1.155\,$eV and $1.136\,$eV, which we relate to the exciton ($X^0_{BL}$) and the trion ($X^T_{BL}$), respectively. The BL shows a smaller trion binding energy of $(19.1\pm 0.1)\,$meV due to a increased screening of the e-h interaction with increasing thickness. 
Our observed values for peaks and binding energies for ML and BL at low temperature are consistent with previous studies \cite{Helmrich,Lezama, Robert,Koirala,Yang}.

\par
\begin{figure}
\centering
\includegraphics[width=0.425\textwidth]{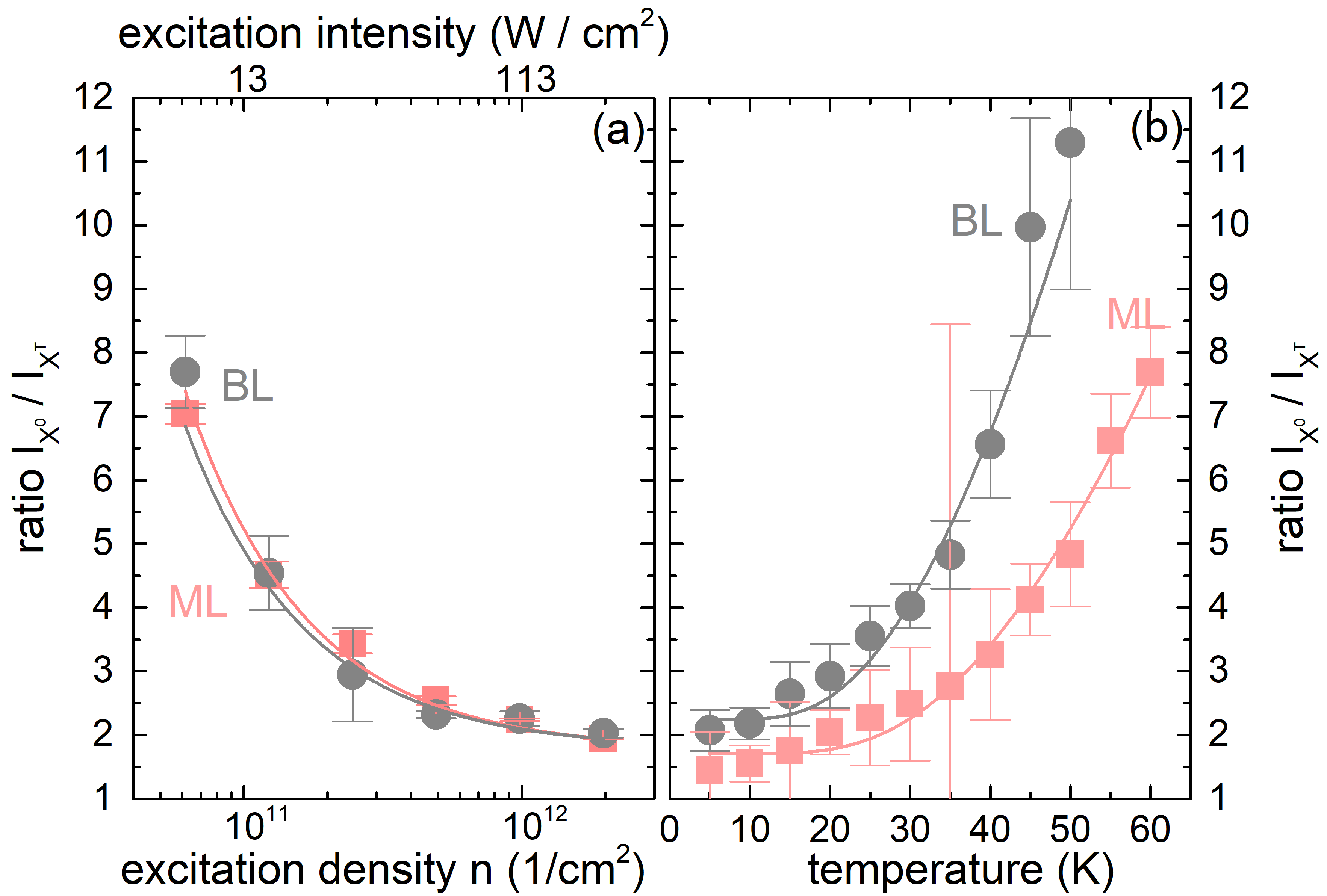}
\caption{(color online) Excitation density $n$ dependent exciton to trion intensity ratio (a) of ML (light red squares) and BL (gray circles). (b) Temperature dependence of ML and BL exciton to trion intensity ratio. Both data are fitted parallel by equation~\ref{2DEG}.}
\label{PowerTrion}
\end{figure}

 Under non-resonant excitation the trion formation is governed by excitons capturing extra electrons from the two dimensional free electron gas of low density \cite{Manassen,Ross}, and its probability thus scales inversely proportional to the excitation density. From PL measurements using the Ti:Sa laser in the mode-locked regime, we extract absolute numbers for the excitation density $n$, see SI. With this number quantitatively specified, the exciton-trion equilibrium can be analyzed using the mass action law. References \onlinecite{Manassen,Ross,Berney,Ron,Robart} predict a dependence on the temperature $T$ and the excitation density $n$ in the equilibrium state of the exciton to trion intensity ratio according to the mass action law of
\begin{equation}
\frac{I_{X^0}}{I_{X^T}} =I_0+ \frac{\tau_{X^T}}{\tau_{X^0}} g_{eff} e^{-E_{\mathrm{bind}}/k_B T}~ \frac{k_BT M_{XT}}{\pi \hbar^2 n}.
\label{2DEG}
\end{equation}
Here $I_0$ is the exciton to trion intensity ratio at zero Kelvin, $\tau_{X^T}/\tau_{X^0}$ is the lifetime ratio of trion to exciton \cite{Matsusue}. $g_{eff}=g_e^2g_{X^0}/g_{X^T}=4$ is the effective spin degeneracy factor, for electrons $g_e$, excitons $g_{X^0}$, and trions $g_{g^T}$ all equal two, assuming a non-degenerate regime \cite{Berney}. $k_B$ is the Boltzmann constant, $M_{XT} = 0.66 ~m_e$ is the mass ratio of exciton to trion \cite{Rasmussen}. We plot the excitation-density and temperature-dependent ratio for $n$ between $6\cdot10^{10}$ and $2\cdot 10^{12}~$cm$^{-2}$ in Fig.~\ref{PowerTrion}~a) and temperatures between 5~K and 60~K in Fig.~\ref{PowerTrion}~b), and simultaneously fit the data by eq.~\ref{2DEG} for ML (light red squares) and BL (gray circles). Both dependencies are well reproduced by an exciton-trion equilibrium described by eq.~\ref{2DEG}. 

\par From the fits, we extract values for the trion to exciton radiative lifetime ratio of $\tau_{X^T}^{ML}/\tau_{X^0}^{ML}=(6 \pm 2)$ and $\tau_{X^T}^{BL}/\tau_{X^0}^{BL} =(3\pm 1)$ for ML and BL, respectively, and thus predict trion radiative lifetimes. Based on time-resolved measurements of the radiative lifetime of the exciton in ML and BL MoTe$_2$ of $3$~ps and $4$~ps \cite{Robert}, we estimate the radiative lifetime of the trion to be $\sim 18$~ps and $\sim12$~ps for ML and BL MoTe$_2$, respectively. The radiative trion lifetime is larger than the exciton lifetime for both ML and BL MoTe$_2$ and corresponds to the ratio of the trion and exciton wave functions \cite{Esser}. Our values extracted for MoTe$_2$ agree very well with the trion and exciton radiative lifetime for ML MoSe$_2$ of $15~$ps to $1.8$~ps measured with a streak camera, which yields a ratio of the radiative lifetimes of ($8\pm1$) \cite{RobertLagarde}.

\par As seen in Fig.~\ref{Gauss}, the lineshapes of the excitonic emission of ML (a) and BL (b) MoTe$_2$ have a Voigtian shape, i.e. a superposition of Gaussian and Lorentzian modifications of the emission line. A Gaussian broadening is indicative of inhomogeneous broadening mechanisms, as it would be the case for a randomly fluctuating local potential, e.g. via defects. Lorentzian (lifetime) broadening mechanisms are homogeneous effects, affecting the whole exciton population equally, such as scattering with phonons.  In Fig.$\,$\ref{Gauss} we extract the full width at half maximum (FWHM) of the Gaussian portion ($\Gamma_G$) of the linewidths for ML and BL at 10~K. Temperature- and excitation intensity- depencence are summarized in the SI.
The Gaussian linewidth of the ML exciton of $7.4\,$meV is comparable to literature values for non-encapsulated MoTe$_2$ \cite{Robert} and encapsulated MoSe$_2$ and MoS$_2$ \cite{Scuri}, thus good quality samples of MoTe$_2$ allow one to study excitons without the influence of hBN. For the ML, $\Gamma_G$ of exciton and trion remains very close. 
This indicates the presence of only one trion species, which at low temperature is an inter-valley trion (cf. Fig.~\ref{Gauss}~a)) with the exciton occupying the K/K' valley and the extra electron residing in the lowest energy level of the conduction band of the K'/K valley \cite{Singh}. For the BL in Fig.~\ref{Gauss}~b), more possible bright trion configurations can be imagined, with the extra charge being confined to either one layer (intra-layer trion) or two layers (inter-layer trion). Values for $\Gamma_G$ differ from exciton to trion by a factor of 1.5 for the BL. As observed previously in WSe$_2$ \cite{YuLiu,Courtade,Singh}, spin degenerate trion states split in energy. Our observations agree with the assumption of a split trion state in the BL with a splitting of  about $J'_{eh}\sim$3-4 meV (cf. SI for calculations).

\begin{figure}
\includegraphics[width=0.42\textwidth]{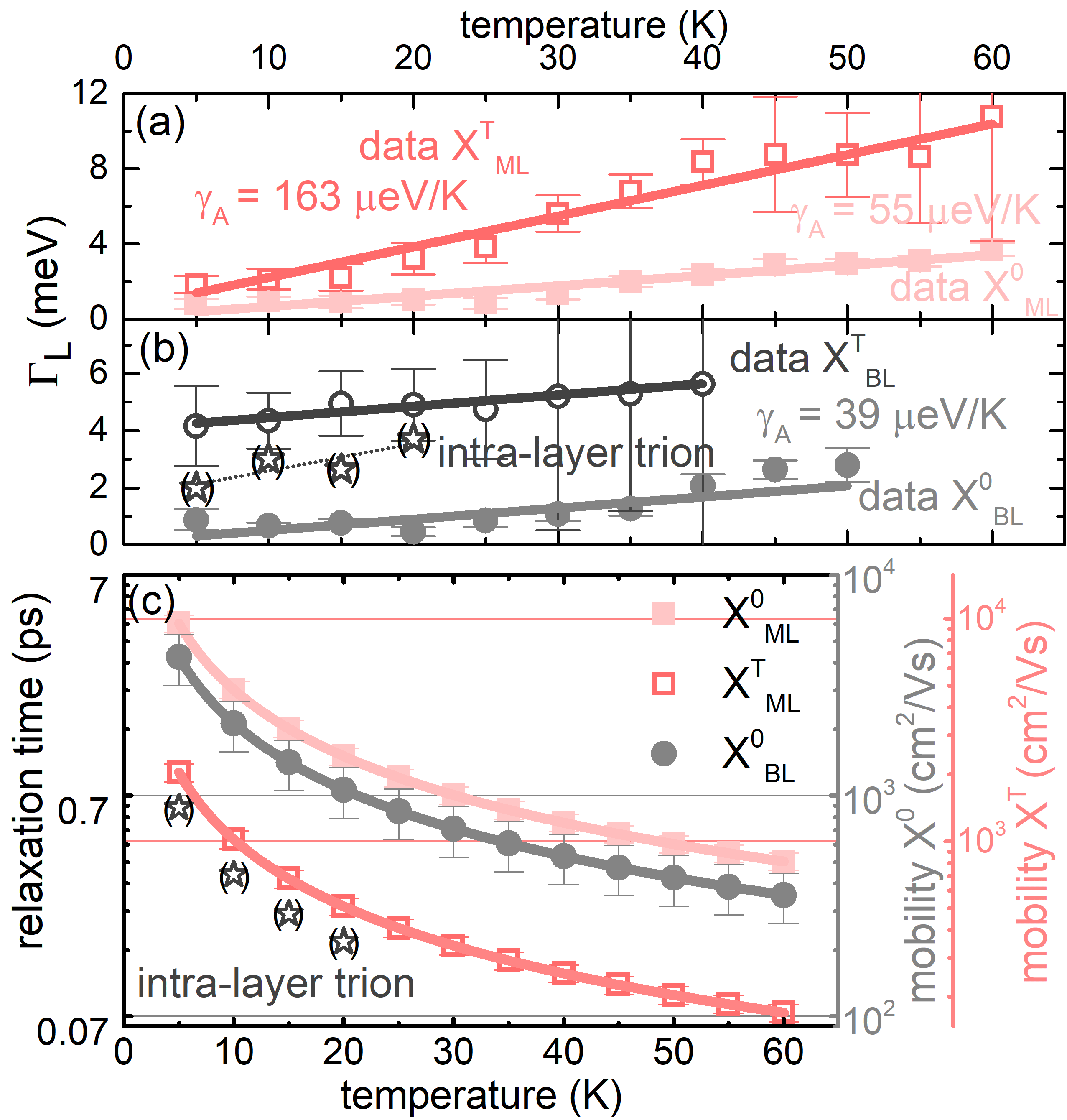}
\caption{(color online) Lorentzian FWHM ($\Gamma_L$) for ML (a) and BL (b) excitonic (closed symbol) and trionic (open symbol) emission versus temperature. The data are fitted by broadening equation. (c) Exciton and trion mobility and relaxation times versus temperature for ML (light red squares) and BL (gray circles) MoTe$_2$.}
\label{Fig2}
\end{figure}

\par The Lorentzian FWHM ($\Gamma_L$) shown in Fig.~\ref{Fig2}~a+b) versus temperature for all quasiparticles increases approximately linearly with temperature in agreement with Ref.~\cite{Koirala}. In case of the BL trion in (b) we extract the Lorentzian FWHM assuming either one trion (open circles) or two trion states where we plot the intra-layer trion $\Gamma_L$ as gray stars. For details we refer to the SI. The slope of the linear function is associated with the quasiparticle-acoustical phonon coupling $\gamma_{A,i}$, $i=X,T$ \cite{Helmrich}. The intra-layer trions display a larger phonon coupling than excitons, which is due their more extended wavefunction and extra charge. This is also confirmed by a vanishing of the trion emission for temperatures higher than 60~K, whereas the exciton emission remains observable up to room temperature.

\par As previously shown in Refs.~\onlinecite{Helmrich} and \onlinecite{Shree}, the temperature-dependence of the homogenous linewidth displayed in Fig.~\ref{Fig2}~a+b) is directly linked to the first order deformation potential $D_X^2=(D_e-D_h)^2$ of the exciton, where $D_{e(h)}$ is the electron (hole) deformation potential. For an estimate of the trion deformation potential we assume that it is a combination of one hole and two electron deformation potentials 
$ D_T =2D_e-D_h$ \cite{Portella-Oberli},
following the argument that the trion emission is a many body effect between excitons and free carriers \cite{Combescot}. Therefore, we can use a similar $\gamma_A$-related ansatz for the trion deformation potential. Taking into account a thickness dependence of the coupling as derived for quantum wells in Ref.~\onlinecite{Basu}, the relation of exciton and trion deformation potential due to acoustical phonons and linewidth broadening is given by
\begin{equation}
\frac{3 k_B T M_i D_i^2}{4 \hbar^2 \rho v_s^2 L}=\gamma_{A,i} T,
\label{Deform}
\end{equation}
where $i=X,T$, $\rho=7.7~ $g~cm$^{-3}$ the density, $v_s$ the sound velocity and $L$ the well thickness. 
The exciton (trion) mass $M_X=m_e^*+m_h^*=1.29~m_e$ ($M_T =2m_e^*+m_h^*=1.93~ m_e$) \cite{Rasmussen} of ML and BL is calculated in a consistent way from the effective electron and hole masses. The published values for the speed of sound in MoTe$_2$ differ from 3.5 km/s (Ref.~\onlinecite{Chen2018}) to 5 km/s (Ref.~\onlinecite{Huang2016}) and are limited to the ML exciton. From atomic force microscope measurements we take the layer thickness of our sample to be 0.66~nm. We obtain upper and lower limits for $D_X$ of 6.3 and 4.4~eV, respectively. These values are very similar to the results obtained for MoS$_2$ and MoSe$_2$ by Shree et al. in Ref.~\onlinecite{Shree}.
We estimate upper and lower limits for $D_T$ of 8.88~eV and 6.22~eV, respectively. From exciton and trion deformation potentials, we can in turn estimate the upper and lower limits of electron (hole) deformation potentials to be $D_e =2.58\,$ and $1.82\,$eV ($D_h =-3.6\,$ and $-2.6\,$eV). These results are of the same order of magnitude as the theoretical estimate in Refs.~\onlinecite{Huang2016,ZhangHuang}. 

\par While the calculation of deformation potentials is complicated by widely differing literature values for effective masses and sound velocities, the acoustic-phonon limited mobility $\mu_{A}$ can be expressed independently of these values (cf. eq.~\ref{mobil}). By combining equations for acoustic phonon limited mobility
\begin{equation}
\mu_{A,i} = \frac{ e \hbar^3 \rho_{2D} v_s^2}{M_i m_{d,i} k_B T D_i^2},
\end{equation}
and deformation potential \cite{Shree,ZhangHuang,Huang2016}, and assuming that the 2D density ($\rho_{2D}$) is the bulk density times half lattice constant $c$=1.39~nm \cite{Shree,Roy}, $\mu_{A,i}$ can be estimated via
\begin{equation}
\mu_{A,i} = \frac{3 e \hbar c }{8 m_{d,i} L}~~ \frac{1}{\gamma_{A,i} T} = \frac{e}{M_i}~ \tau_{k,i}.
\label{mobil}
\end{equation}
The mobility is directly linked to the relaxation time for exciton and trion acoustic phonon scattering $\tau_{k,i}$ \cite{Kaasbjerg2}. Here, $e$ is the elementary charge and $m_d$ is the quasiparticle density of state mass \cite{ZhangHuang}. The resulting mobilities and relaxation times are plotted in Fig.~\ref{Fig2}~c) for exciton and trion of the ML (light red squares) and the BL (dark gray symbols). Note that the values for the relaxation time differ from exciton to trions due to the varying effective quasiparticle mass $M_i$. The mobility scales inversely with temperature, reaching extremely high values at low temperatures. At 5~K, we estimate about 6000~cm$^2$/Vs for the ML exciton, closely approached by 4300~cm$^2$/Vs for the BL exciton. The relaxation times and mobilities of 2000 and 1400~cm$^2$/Vs for the ML and BL intra-layer trion are smaller than for the exciton. This disagrees with our previous results on the radiative lifetime ratio of trion and exciton, cf. Fig.~\ref{PowerTrion}~a+b), hence the mobilities are diffusion-limited rather than lifetime-limited. The BL shows an even smaller radiative lifetime ratio in agreement with the smaller $\gamma_A$ exciton to trion ratio for BL. 
Note that the considerations above yield a mobility only limited by scattering with acoustical phonons. This neglects other processes such as impurity scattering or electron scattering, which can be summed up via the Mathissen rule \cite{Huang2016}.

\begin{figure}
\includegraphics[width=0.42\textwidth]{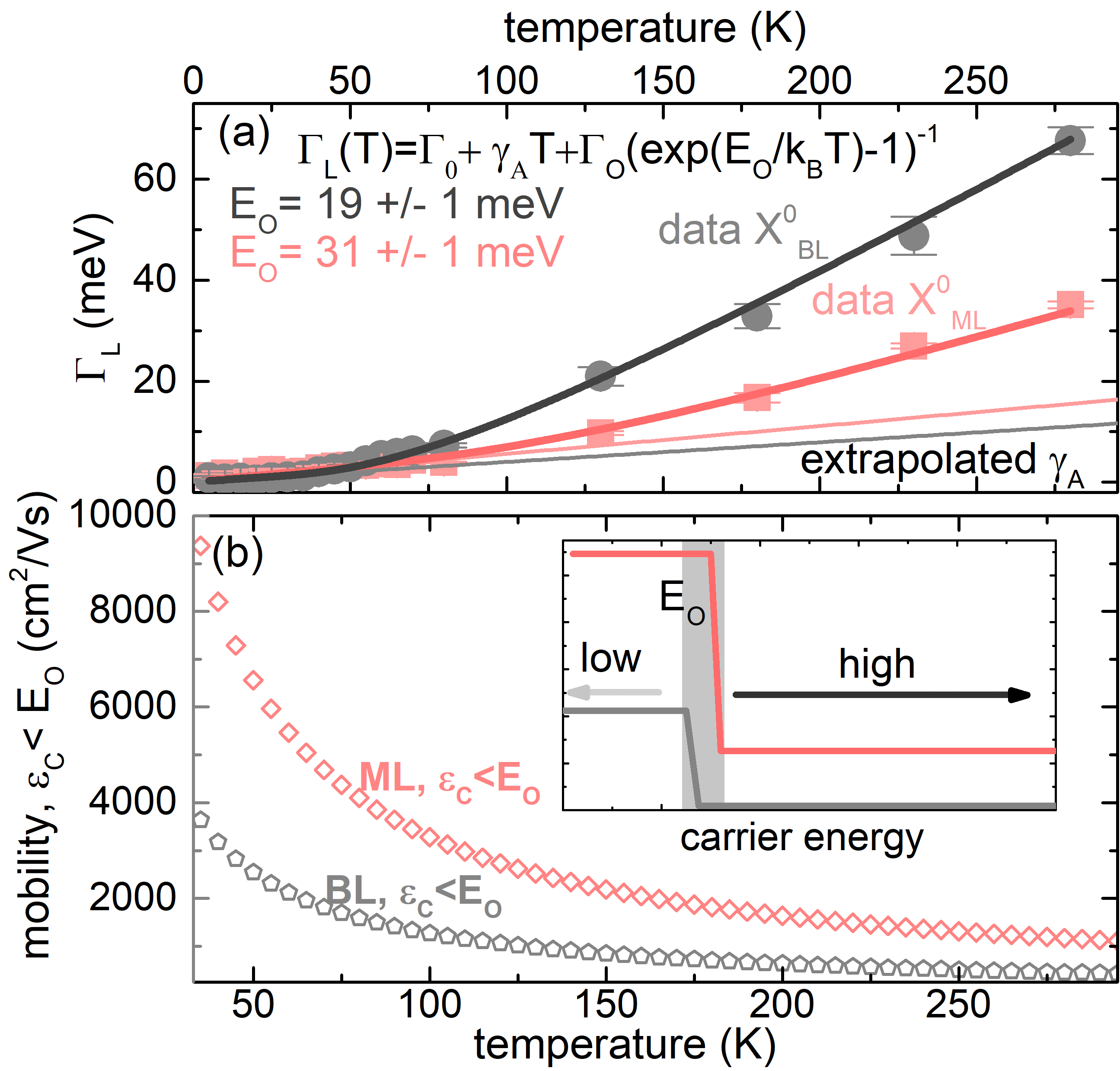}
\caption{(color online) (a) Lorentzian FWHM ($\Gamma_L$) for ML (light red squares) and BL (gray circles) excitonic emission versus temperature. (b) Exciton optical-phonon-limited mobility for carrier energies $\epsilon_K$ below phonon energy $E_O$ versus temperature.}
\label{Fig4}
\end{figure}

\par Apart from the acoustical phonons discussed above, optical phonons play an important role especially at higher temperatures. Due to small optical phonon energies in MoTe$_2$ 
\cite{GuoYang} optical phonons are expected to limit the mobility early on \cite{Huang2016}. The total phonon-limited mobility $\mu_P$ is then given by $1/\mu_P =1/\mu_A +1/\mu_O $, with $\mu_O$ the optical-phonon-limited mobility. We investigate the optical phonon limited mobility by analyzing the Lorentzian linewidth broadening at higher temperatures. 
In Fig.~\ref{Fig4}~a), we plot $\Gamma_L$ over a larger temperature range than above. From these data, we can extract the energy of the optical phonons $E_O$ by fitting the data with $\Gamma_L (T) = \Gamma_0 +\gamma_A T + \Gamma_O [\exp(E_O/k_B T)-1]^{-1}$ for ML and BL MoTe$_2$. Further fitting parameters are discussed in the SI.
The resulting values for $E_O$ are $(31\pm1)\,$meV and $(19\pm1)\,$meV for ML and BL, respectively. Comparing $E_O$ to the calculated phonon dispersions in Ref.~\onlinecite{GuoYang}, mostly longitudinal optical (LO) phonons contribute to the exciton phonon coupling, mainly polar and homopolar LO phonons in case of the ML and polar LO and transversal optical (TO) phonons in case of the BL. Under the assumption that (i) both LO modes (polar+ homopolar) correspond to a zero order deformation potential of the wavevector as predicted by Ref.~\onlinecite{Kaasbjerg} and (ii) TO phonon modes correspond to a first order deformation potential $D^1$, however, $D^1$ is much smaller than $D^0$ for the low and moderate energies we use in this contribution, we can restrict our estimate to scattering described by zero order deformation potential $D^0$ theory. Thus, the optical-phonon-limited mobility can be estimated by \cite{Kaasbjerg}
\begin{equation}
\mu_O = \frac{2 e \hbar \rho_m E_O}{M^2 (D^0)^2} \left[ 1+e^{\frac{E_O}{k_B T}} \Theta(\epsilon_K-E_O)\right]^{-1} N_0^{-1}, 
\end{equation}
where $D^0 = 1.34 \cdot 10^8\, $eV cm$^{-1}$ is calculated in Ref.~\onlinecite{Huang2016}, $N_0\sim k_BT/E_O$ is the Bose-Einstein distribution in first approximation, and $\Theta(\epsilon_K-E_O)$ is the Heaviside step function \cite{Kaasbjerg}, which assures that only excitons with a certain energy can emit a phonon. Precisely, the equation above is designed for fermionic particles describing the mobility of electrons. In MoTe$_2$ electrons and holes have comparable masses \cite{Rasmussen}, thus, a comparable curvature of the band structure and group velocity. We thus assume that exciton as localized electron-hole pair moving in the lattice experiences comparable movement limitations. 
We calculate the mobility for a carrier energy $\epsilon_K$ 
below the phonon energy $E_O$, and plot this in Fig.~\ref{Fig4}~b) versus temperature.
The calculations are limited to excitons here, as the trion signature disappears at higher temperatures, and the optical phonons interaction becomes significant only at temperatures above 40~K. For lower carrier temperatures, we get an extremely high optical-phonon-limited mobility for $\epsilon_K< E_O$ indicating that such excitons are not impaired by optical phonons. For high temperatures (room temperature) optical phonons limit the mobility of these low-energy carriers strongly. 
We can estimate an upper 
value of $1100\,$cm$^2$/Vs  and $250\,$cm$^2$/Vs 
for the optical-phonon-limited mobility in ML and BL 
at room temperature. Compared to other TMDCs, the ML displays a high mobility in the room-temperature limit \cite{Kaasbjerg,Kaasbjerg2,Huang2016,ZhangHuang}, 
whereas the BL shows values comparable to the ML mobility of other members of the TMDC family. The deformation potential, acoustical phonon limited mobility at 5~K and optical phonon limited mobility at room temperature are summed up in table~\ref{tab:FittingConstants}.

\begin{table}
\caption{Deformation potentials $D^{max}$ and $D^{min}$, acoustical  $\mu_A$ and optical $\mu_O$ phonon limited mobility for ML exciton, trion, electron, and hole and BL exciton.}
\label{tab:FittingConstants}
\centering
\begin{tabular}{lcccc}
\hline
\multicolumn{5}{l}{\textbf{Deformation potential and mobility ML}} \\
\hline
& $D^{max}$~(eV) & $D^{min}$~(eV) & $\mu_A$~(cm$^2/$Vs) & $ \mu_O$~(cm$^2/$Vs)  \\
X & 6.3 & 4.4 & 6000 & 1100\\
T & 8.9 & 6.2 &  2000 &  \\
e & 2.6 & 1.8 & 61000 &  \\
h & -3.6 & -2.6 & 29000 &   \\
\hline
\multicolumn{5}{l}{\textbf{Deformation potential and mobility BL}} \\
\hline
& $D^{max}$~(eV) & $D^{min}$~(eV) & $\mu_A$~(cm$^2/$Vs) & $ \mu_O $~(cm$^2/$Vs) \\
X & 7.5 & 5.2 & 4300 & 250 \\
T$_{intra}$ & 9.6 & 6.7 & 1400 &  \\
\hline
\end{tabular}

\end{table}

\par In conclusion, we have investigated the lineshape of the photoluminescence of mono- and bilayer MoTe$_2$ over a wide range of temperatures up to room temperature. Both show a bright photoluminescence with two distinct peaks. The lower energy peak in the emission of the bilayer, which has previously only tentatively been assigned to a negatively charged trion, shows a temperature and excitation intensity dependence in excellent agreement with this assumption. The peak is possibly originating from both an intra- and an inter-layer trion split by 3-4 meV in energy. From the temperature-dependent lineshapes of the luminescence, we derived deformation potentials and acoustical and optical phonon-limited carrier and quasiparticle mobilities in the MoTe$_2$ system. We find a very high mobility in both ML and BL, making MoTe$_2$ the most suited TMDC material for opto-electronics applications.

\textbf{Acknowledgments}
We thank Robert Schneider from the group of Rudolf Bratschitsch at Westf\"alische Wilhelms University M\"unster for exfoliating the excellent quality sample.
Funding for this research was provided by Deutsche Forschungsgemeinschaft via the GRK 1558, the CRC 787, grant No. AC290/2-1 and AC290/2-2 (A. W. A.).

\nocite{*}
\bibliography{MobilityHelmrich}% Produces the bibliography via BibTeX.

\end{document}